\documentstyle[psfig,draft,aps]{revtex}
\begin{document}
\draft
\title{
%{\begin{flushright}{\footnotesize ZU-TH 9/00}\\
%{\footnotesize YU-PP-I/E-KM-6-00}
%\end{flushright}}
The strange quark mass from flavor breaking in 
hadronic $\tau$ decays}
\author{Joachim Kambor\thanks{e-mail: kambor@physik.unizh.ch}}
\address{Institut f\"ur Theor. Physik, Univ. Z\"urich,
CH-8057 Z\"urich, Switzerland}
\author{Kim Maltman\thanks{e-mail: maltman@fewbody.phys.yorku.ca}}
\address{Department of Mathematics and Statistics, York University, \\
4700 Keele St., Toronto, Ontario, Canada M3J 1P3 \\ and }
\address{Special Research Centre for the Subatomic Structure of Matter, \\
University of Adelaide, Australia 5005.}
\maketitle
\begin{abstract}
The strange quark mass is extracted from a finite energy
sum rule (FESR) analysis of the flavor-breaking difference of 
light-light and
light-strange quark vector-plus-axial-vector correlators, using spectral
functions determined from hadronic $\tau$ decay data.
We point out 
problems for existing FESR treatments associated with potentially
slow convergence of the perturbative series 
for the mass-dependent terms in the OPE 
over certain parts of the FESR contour, and
show how to construct alternate
weight choices which not only cure this problem,
but also (1) considerably improve the convergence of the integrated
perturbative series, (2) strongly suppress contributions from
the region of $s$ values where the errors on the strange
current spectral function are still large and (3) essentially
completely remove uncertainties associated with the
subtraction of longitudinal contributions to the experimental
decay distributions.
The result is an extraction of $m_s$
with statistical errors comparable
to those associated with
the current experimental uncertainties in the determination of the
CKM angle, $V_{us}$.  We find
$m_s(1\ {\rm GeV})=158.6\pm 18.7\pm 16.3\pm 13.3\ {\rm MeV}$ (where the first 
error is statistical, the second due to that on
$V_{us}$, and the third theoretical).
\end{abstract}
\section{Introduction}
The light quark masses, $m_s$, $m_u+m_d$, are among the least well
determined of the fundamental parameters of the Standard Model
and, as such, have been the subject of much recent attention,
in both the QCD sum 
rule~\cite{bpr,jm,cps,n38,cfnp,bgm,km38,dps,kmss,ck93,kmtpr,pp98,cdh,pp99,ALEPHstrange,n3s,nrev}
and lattice~\cite{lat1,lat2,lat3,latticerecent} 
communities.

Recent attempts to extract $m_u+m_d$ and $m_s$
via sum rule analyses of, in the former case,
the light quark ($ud$) pseudoscalar
correlator\cite{bpr}, and in the latter case, the
light-strange ($us$) scalar\cite{jm,cps,cfnp,kmss}
or pseudoscalar\cite{dps} correlators, 
suffer from the problem that the relevant spectral
functions are not fully determined experimentally in the
region required for the analyses.

Analyses based on vector current correlators involving
various pieces of the light quark electromagnetic
(EM) current suffer from analogous problems.
In the case of Narison's sum rule based on the difference
of the flavor $33$ (isovector) and $88$ (hypercharge, or
isoscalar) correlators~\cite{n38}, the
$G$-parity-based
identification of the $33$ and $88$ contributions
to the EM hadroproduction cross-section,
which would allow the difference of $33$ and $88$ spectral functions
to be determined from experimental data,
is valid only in the absence of isospin breaking (IB).
The high degree of cancellation (to the level of $10-15\%$) 
between the $33$ and $88$ spectral integrals
makes the analysis rather sensitive to the neglect of 
IB~\cite{km38}.  This sensitivity is compounded by the fact that a
sum rule determination of the corrections required
to remove the $38$ contributions from the experimental data
shows that, for reasons which are easily understood~\cite{km38}, 
the dominant corrections, associated with
the $\omega$ contribution to the nominal $88$ spectral
function\cite{km38,kmcew98}, are larger
than one would naively expect.\begin{footnote}{The central value 
$m_s (1\ {\rm GeV})=176\ {\rm MeV}$~\cite{n3s},
obtained neglecting IB corrections, is reduced to
$146\ {\rm MeV}$ when one applies the IB corrections
obtained in the sum rule analysis of Ref.~\cite{kmcew98}.}\end{footnote}
The necessity of determining the IB corrections
theoretically thus prevents one from working with a sum
rule whose spectral side is determined solely by experimental data.

A similar problem exists
for the sum rule based on the
difference of $33$ and $ss$ vector current
correlators~\cite{n3s}, since the portion of 
the EM hadroproduction cross-section associated
with the $ss$ part of the EM spectral function
is not an experimental observable.
In Ref.~\cite{n3s}, it is assumed to be given by
the cross-section for the production of the various
$\phi$ resonances.  This approximation, while
no doubt a reasonable one, 
is exactly valid
only if both (1) the Zweig rule
is $100\%$ satisfied and (2) the $\phi$ resonances 
are all pure flavor $\bar{s}s$ states.  The close
cancellation (to the $\sim 15\%$ level)
between the $33$ and $ss$ spectral integrals again makes
the analysis sensitive to even small (few $\%$) Zweig rule violations
(ZRV).  To illustrate this sensitivity,
let us take the deviation from ideal mixing in the vector meson sector
as a measure of the natural scale of 
ZRV,~\begin{footnote}{From Ref.~\cite{pdg98} 
one has that the vector meson mixing angle
is either $36^o$ or $39^o$, depending on whether
one uses the linear or quadratic mass formula.}\end{footnote}
and consider a scenario in which ZRV
occurs dominantly in the mass matrix and not in the
vacuum-to-vector-meson matrix elements of 
the vector currents.  The strange (light) quark part of the EM current
then couples only to the strange (light) part of any given resonance.
If the flavor content of a given $\phi$ resonance is
$\alpha \bar{s}s +\beta (\bar{u}u+\bar{d}d)/\sqrt{2}$
(with $\alpha\simeq 1$ and $\beta$ small), 
the ratio of the square of the full EM $\phi$ decay
constant to that of the decay constant describing the coupling
only to the $ss$ part of the EM current is then
$\simeq 1-\sqrt{2}\beta /\alpha$.  For either the linear
or quadratic versions of mixing this ratio is less than $1$;
including ZRV corrections will thus increase the
$ss$ spectral function and hence
lower the extracted value of $m_s$.  Taking, to be specific, the
case that the radius of
the circular part of the FESR contour is $(1.6\ {\rm GeV})^2$,
we find that, using an identical method of analysis and
identical higher dimensional condensate values to those
employed in Ref.~\cite{n3s} (and including,
for completeness, the small IB isovector contribution
to the $\phi (1020)$ EM decay constant determined in Ref.~\cite{kmcew98}),
the central value of $m_s(1 \ {\rm GeV})$ obtained ignoring
IB and ZRV~\cite{n3s} ($196\ {\rm MeV}$) is lowered to $177\ {\rm MeV}$
($108\ {\rm MeV}$) for the linear (quadratic) cases,
respectively.  We stress that the point of this exercise is not
to attempt a realistic estimate of ZRV corrections
but rather to point out that,
given the scale at which such violations are already {\it known}
to occur, the uncertainties in the extraction of
$m_s$ associated with the neglect of ZRV
are large, and, moreover, cannot be
significantly reduced without a major improvement in our theoretical
understanding of the precise nature and magnitude of ZRV.\begin{footnote}
{In Ref.~\cite{n3s}, the agreement of the 
$33$-$88$ and $33$-$ss$ determinations of $m_s$
obtained ignoring IB and ZRV, respectively, was taken as
evidence against the size of the IB corrections
obtained in Ref.~\cite{kmcew98}.  Note, however, that
(1) within errors, the latter result is compatible with 
either the IB-corrected or uncorrected $33$-$88$
determination, and (2) two inverse moment
sum rule determinations of the
$6^{th}$ order chiral low-energy constant, $Q$, one based
on the $33$-$88$~\cite{gk96},
and one on the $\bar{s}u$-$33$ correlator difference~\cite{dk99},
are brought into almost perfect
agreement once the IB corrections of Ref.~\cite{kmcew98} are applied
to the former analysis.}\end{footnote}

In light of the fact that, in each of the analyses above,
it is not possible to work with sum rules for which
the hadronic spectral function is 
determined entirely by experimental
data, we will, in this paper, instead construct
finite energy sum rules (FESR's)
based on the flavor-breaking difference
between the sum of the $ud$ vector and axial
vector correlators and the corresponding sum of
$us$ correlators, for which, up to $s=m_\tau^2$, the
spectral function can be taken from experimental
hadronic $\tau$ decay data~\cite{ALEPHlight,ALEPHstrange}.
The rest of the paper is organized as follows.
In Section II we provide a brief review, and discuss the
practical difficulties to be overcome in arriving at a
reliable implementation of this approach.
In Section III we describe a construction which leads to
FESR's which successfully overcome 
these difficulties, and in Section IV we give numerical details
and discuss our results.

\section{Flavor-Breaking Sum Rules Involving Hadronic $\tau$
Decay Data}
For a general correlator, $\Pi (s)$, with a cut 
beginning at $s=s_{th}$ and running along the timelike
real axis, one obtains from Cauchy's theorem,
defining the spectral function, as usual,
by $\rho\equiv {\rm Im}\, \Pi /\pi$,
the general FESR relation
\begin{equation}
\int_{s_{th}}^{s_0}\, ds\, \rho (s) \, w(s)\, = {\frac{-1}{2\pi i}}\,
\oint_{\vert s\vert =s_0}\, ds\, \Pi (s) w(s) \,
\label{basicfesr}
\end{equation}
where $w(s)$ is any function analytic
in the region of the contour, $C$, consisting of the union of
the circle of radius $s_0$ in the complex $s$-plane and
the lines above and below the physical cut, running from 
$s_{th}$ to $s_0$.

As is well known, the ratios of $ud$ and $us$
inclusive hadronic $\tau$ decay widths to the $\tau$ electronic
decay width,
\begin{equation}
R^{ij}_\tau \equiv \frac{ \Gamma [\tau^- \rightarrow \nu_\tau
\, {\rm hadrons_{ij}}\, (\gamma)]}{ \Gamma [\tau^- \rightarrow
\nu_\tau e^- {\bar \nu}_e (\gamma)] } ,
\label{one}\end{equation}
where $(\gamma )$ indicates additional photons or lepton pairs,
and $ij=ud,us$ labels the flavors of the 
relevant portion of the hadronic weak current, can be expressed
as weighted integrals over the relevant
spectral functions.  Eq~(\ref{basicfesr}) then allows these ratios
to be recast into a form appropriate for the 
use of techniques based on the OPE
and perturbative QCD~\cite{tau1,tau2,bnp,ledp92,pichrev}.
Letting $J^\mu_{ij;V,A}$ be the usual vector and axial vector
currents with flavor content $ij$, and 
defining the scalar $J=0,1$ parts of
the corresponding correlators by
\begin{eqnarray}
i \int d^4x \, e^{iq\cdot x} &&
\langle 0\vert T\Bigl(
J^{\mu}_{ij;V,A}(x) J^{\nu}_{ij;V,A}(0)^\dagger 
\Bigr)\vert 0\rangle \, \nonumber \\
&&\qquad\qquad\equiv \left( -g^{\mu\nu} q^2 + q^{\mu} q^{\nu}\right) \, 
\Pi_{ij;V,A}^{(1)}(q^2)
 + q^{\mu} q^{\nu} \, \Pi_{ij;V,A}^{(0)}(q^2)\, ,
\end{eqnarray}
one has 
\begin{eqnarray}
R^{ij}_\tau   &=&  
12 \pi^2 S_{EW}\, \vert V_{ij}\vert^2 
\int^{m_\tau^2}_0 {\frac{ds} {m_\tau^2 }} \,
\left( 1-{\frac{s}{m_\tau^2}}\right)^2 
\left[ \left( 1 + 2 {\frac{s}{m_\tau^2}}\right) 
\, \rho^{(1)}_{ij}(s) 
+ \, \rho^{(0)}_{ij}(s) \right] \, 
\nonumber \\
&=& 6 \pi S_{EW}\,  \vert V_{ij}\vert^2 
i\, \oint_{|s|=m_\tau^2} {\frac{ds}{ m_\tau^2}}\,
\left( 1- {\frac{s}{ m_\tau^2}}\right)^2 \left[ \left( 
1 + 2 {\frac{s}{m_\tau^2}}\right) \Pi^{(0+1)}_{ij}(s)
- 2 {\frac{s}{ m_\tau^2}}\, \Pi^{(0)}_{ij}(s) \right] \, ,
\label{taufesr}
\end{eqnarray}
where $\Pi^{(J)}_{ij}\equiv\Pi^{(J)}_{ij;V}+\Pi^{(J)}_{ij;A}$,
$\rho^{(J)}_{ij} (s)$ are the corresponding spectral functions,
$S_{EW}=1.0194$ represents the leading electroweak 
corrections\cite{ms88}, and $V_{ij}$ are the usual CKM matrix elements.
Since $m_\tau^2\sim 3\ {\rm GeV}^2$, the second expression 
in Eq.~(\ref{taufesr}) 
is amenable to evaluation using the OPE.  
Dividing both the hadronic and OPE 
expressions by $\vert V_{ij}\vert^2$, and taking the difference
of the $ij=ud$ and $us$ cases, 
one arrives at a flavor-breaking FESR
\begin{eqnarray}
&&\int^1_0 {dy} \,
\left( w_{L+T}(y) 
\Delta\rho^{(0+1)}(s)\, + \, w_L(y)\Delta\rho^{(0)}(s)
\right) \nonumber \\
&&\qquad\qquad\qquad\qquad =
{\frac{-1}{2\pi i}}\oint_{\vert y\vert =1}\, dy\,
\left( w_{L+T}(y)\Delta\Pi^{(0+1)}(s)+w_L(y)
\Delta\Pi^{(0)}(s)\right)
\label{fbfesr}
\end{eqnarray}
where $y\equiv s/m_\tau^2$, $\Delta\Pi^{(J)}\equiv
\Pi^{(J)}_{ud}-\Pi^{(J)}_{us}$,
$\Delta\rho^{(J)}\equiv\rho^{(J)}_{ud}-\rho^{(J)}_{us}$,
and $w_{L+T}$, $w_L$ refer to the longitudinal-plus-transverse
($(J=0)+(J=1)$, or ``$L+T$'')
and ``longitudinal'' ($(J=0)$) kinematic
weights $w_{L+T}(y)\equiv \left(1-y\right)^2\left( 1+2y\right)$
and $w_L(y)=-2y\left(1-y\right)^2$, respectively.
The mass-independent ($D=0$) piece of the correlator difference
$\Delta\Pi^{(J)}$ on the OPE side of the sum rule Eq.~(\ref{fbfesr})
of course vanishes by construction.  
In the limit that we neglect $m^2_{u,d}$ and $\alpha_s m_{u,d}m_s$ 
relative to $m_s^2$, moreover, the $D=2$ terms in the OPE representation of 
$\Pi_{V+A;ij}^{(J)}$ become simply proportional to $m_s^2$.
Were the OPE representations
of both the $L+T$ and longitudinal contributions above
to be well converged at scale $m_\tau^2$,
Eq.~(\ref{fbfesr}) would thus allow a determination
of $m_s$ in terms of the difference of experimental non-strange and
strange decay number distributions.  

The perturbative series for the 
integrated $D=2$ longitudinal contribution in Eq.~(\ref{fbfesr}),
however, turns out not to be convergent at the scale
$s_0=m_\tau^2$~\cite{kmtpr,pp98},
creating a serious problem for the analysis in the absence of
an experimental separation of transverse and longitudinal spectral
contributions.  This separation is straightforward
at low $s$ but experimentally problematic above
$1\ {\rm GeV}^2$.\begin{footnote}
{In Ref.~\cite{kmtpr}, an attempt was made to circumvent this
problem by assuming the validity, even in the
region of non-convergence,
of a relation between
the integrated longitudinal OPE vector and axial vector $D=2$ 
contributions valid
in the region of convergence of the OPE representations of both.
If true, this would allow the longitudinal strange axial integral
to be obtained from the longitudinal strange vector integral.  The
latter can be obtained using 
the model strange scalar spectral function of Ref.~\cite{cfnp}.
Using appropriately-weighted FESR's for
the strange pseudoscalar channel, we have now been able
to test this
assumption, and demonstrate that it is, in fact, incorrect.}
\end{footnote}
Our inability to treat
the OPE representation of the longitudinal contributions
in a reliable manner thus
creates difficult-to-quantify uncertainties
for any FESR involving significant 
longitudinal spectral contributions.
Existing analyses
are included in this category since, for example,
the central value for the difference of non-strange and strange
spectral integrals
from the analysis of
Refs.~\cite{cdh,ALEPHstrange},
\begin{equation}
\Delta^{00}\equiv {\frac{R^{ud}_\tau}{\vert V_{ud}\vert^2}}-
{\frac{R^{us}_\tau}{\vert V_{us}\vert^2}}=0.394\pm 0.137\, ,
\label{ALEPHdel00}
\end{equation}
corresponds to $L+T$, longitudinal and higher dimension
condensate contributions which are $0.184$, $0.155$ and
$0.055$, respectively.  

Another practical problem is the close cancellation
between the rescaled $us$ and $ud$ spectral integrals for the
sum rules above, based on the kinematic weights, $w_{L+T}$ and $w_L$.
In the analysis of Refs.~\cite{cdh,ALEPHstrange}, for example, 
the cancellation is to the $\sim 10\%$ level,
making the results very sensitive
to both small variations in the input parameters and 
the sizeable experimental errors ($\sim 20-30\%$)
on the strange decay number distribution
above the $K^*$ region.
Two features of
the analysis of Refs.~\cite{cdh,ALEPHstrange} illustrate the former
sensitivity.  First, Refs.~\cite{cdh,ALEPHstrange}
employ $\vert V_{us}\vert =0.2218\pm 0.0016$,
{\it c.f.} the PDG98~\cite{pdg98}
value $0.2196\pm 0.0023$.  Though compatible within errors,
the squares of the two central values differ by $\sim 2\%$;
use of the PDG98 value {\it decreases} the flavor-breaking
difference, $\Delta^{00}$, by $17\%$.  Since one cannot
reliably employ the OPE representation of the longitudinal
contributions, moreover, 
the longitudinal spectral contribution (which is dominated, at
the $\sim 80\%$ level, by the $K$ pole term)
must be subtracted; the
shift in the inferred $L+T$ contribution
(used to determine $m_s$)
is thus even larger ($36\%$).  Similarly,
use of the PDG98 value $f_K=113.0\pm 1.0\ {\rm MeV}$ in place
of the ALEPH determination, $f_K= 111.5\pm 2.5\ {\rm MeV}$
lowers the inferred
$L+T$ contribution to $\Delta^{00}$ by a further $12\%$.  The
combined impact on the central value for $m_s$ is thus
extremely large, though the two central values are, of course,
compatible within the (large) errors quoted in 
Refs.~\cite{cdh,ALEPHstrange}.
The {\it relative} size of the residual statistical
errors as a fraction of the resulting $\Delta^{00}$
is, of course, also significantly increased by such
a decrease in $\Delta^{00}$.
It is thus highly desirable to choose,
in place of the kinematic weights, weights which produce a less close
cancellation between the $ud$ and $us$ spectral
integrals.  The easiest way to accomplish this goal is to choose
weight functions which
fall off more rapidly through the region of
the excited strange resonances.
This has the happy consequence of also suppressing 
contributions from the region where both the 
errors on the strange spectral distribution are
large and the transverse/longitudinal separation is experimentally
difficult.

The final difficulty to be dealt with is theoretical.
Suppose we are able to solve the
longitudinal/transverse separation problem, and
thus work with FESR's
involving only 
the $L+T$ part of the flavour breaking difference,
\begin{equation}
\Pi(q^2) \equiv \Pi_{ud,V+A}^{(1+0)}-\Pi_{us,V+A}^{(1+0)}\ .
\label{defPi}
\end{equation}
The leading ($D=2$) $m_s$-dependent terms
in the OPE representation of $\Pi$
are~\cite{ck93}
\begin{eqnarray}
\left[ \Pi (Q^2)\right]_{D=2}&=& -{\frac{3}{2\pi}}{\frac{
m^2_s(Q^2)}{Q^2}}\left[ 1+{\frac{7}{3}} a(Q^2)
+(19.9332 ) a(Q^2)^2+\cdots \right] \nonumber \\
&\equiv&  -{\frac{3}{2\pi}}{\frac{
m^2_s(Q^2)}{Q^2}}\sum_{k=0}g_k\, a(Q^2)^k \ ,
\label{d2trans}\end{eqnarray}
with $a(Q^2) =\alpha_s(Q^2)/\pi$
and $m_s(Q^2)$ the running coupling and running strange quark mass, both
at scale $\mu^2=Q^2=-s$, in the $\overline{MS}$ scheme.
The ratio of ${\cal O}(a)$
and ${\cal O}(a^2)$ coefficients in Eq.~(\ref{d2trans}) is 
rather large ($8.5$), signalling potentially slow
convergence (with $\alpha_s(m_\tau^2)
=0.334$~\cite{ALEPHlight}, the ratio of the
${\cal O}(a^2)$ 
and ${\cal O}(a)$ terms is $0.90$
at $\mu^2=m_\tau^2$, and $>1$ for $\mu^2$ below $\sim 2.2\ {\rm GeV}$.)
In recent analyses~\cite{cdh,pp99,ALEPHstrange}, this potential problem
is brought under (apparent) control using the
method of ``contour improvement''~\cite{ledp92}.
In this method, the logarithms in $\Pi$ are first summed
(as has already been done in Eq.~(\ref{d2trans})) by choosing the
renormalization scale equal to $Q^2$ at each point on the
circle $\vert s\vert =s_0$.  The integrals
\begin{equation}
A^{[w_{L+T}]}_k(s_0)={\frac{-1}{2\pi i}}\oint_{\vert s\vert=s_0}\, ds\, 
\left[ {\frac{m(Q^2)^2}{Q^2}}\right] \, a(Q^2)^k\, w_{L+T}(y);\qquad
y=s/s_0
\label{contourorder}
\end{equation}
are then evaluated numerically, using the known $4$-loop forms
for the running mass and coupling.  The OPE side of the $L+T$
part of the conventional $\tau$ decay sum rule then reduces
to a linear combination of the $A^{[w_{L+T}]}_k(m_\tau^2)$, $k=0,1,2$,
with the index $k$ giving the ``contour-improved order''.
Both the convergence and the residual scale dependence of the resulting 
truncated series are significantly improved by
this procedure~\cite{pp98,pp99}.  
Since, relative
to an expansion in terms of $a(\mu^2)$, for some fixed scale $\mu^2$,
contour improvement represents a resummation of the perturbative
series, it is possible that this improvement is physically
meaningful.  

Unfortunately, it turns out that the apparent
improvement is not a general one, but rather
the result of an accidental suppression of the $k=2$ integral.
To see this, 
let us, for illustrative purposes, imagine that the unknown
coefficients, $g_k$, for $k\geq 3$, in Eq.~(\ref{d2trans})
grow geometrically, i.e., $g_k = (19.9332)
\left[ {\frac{19.9332}{7/3}}\right]^{k-2}, k\geq 3$.{\begin{footnote}
{Note that Refs.~\cite{cdh,pp99,ALEPHstrange} employ a form of
the $L+T$ FESR in which the OPE integral has been partially integrated
once in order to re-express it in terms of the difference of
$L+T$ $ud$ and $us$ Adler functions.  The contour-improved
series for the Adler function version differs term-by-term
from that based on the direct correlator difference.  Though
the agreement of the sums of the two versions to second order
is excellent, the reader should bear in
mind that the relative size of the terms of different
order is not the same in the two cases.}
\end{footnote}}
We then evaluate $A^{[w_{L+T}^N]}_k(s_0)$ for $k=0,\cdots ,10$ 
and $s_0=m_\tau^2$, where $w_{L+T}^N(y)=w_{L+T}(y)[1-y]^N$, $N=0,1,2$,
are the ``spectral weights'' employed in the analyses of 
Refs.~\cite{cdh,pp99,ALEPHstrange}.
The results of this exercise, rescaled in each case 
by the corresponding $k=0$ value, are displayed in Table I.
In columns 2-4 we see the apparently favorable convergence of
the $k=0,1,2$ terms already discussed.  The results of the
remaining columns, however, 
show that the smallness of the $k=2$ term is not the result
of a favorable resummation (which would lead also
to improved convergence for the remainder of the series)
but rather a consequence of the fact
that $A^{[w_{L+T}^N]}_k(m_\tau^2)$ has a zero
as a function of $k$ rather close to $k=2$.  The magnitudes of the
$k\geq 3$ terms are such that 
truncation of the series at $k=2$ would produce
a significant theoretical error, one much larger in
magnitude than the size of the $k=2$ term.{\begin{footnote}
{One should bear in mind that, were one to work with the
Adler function version of the $L+T$ FESR, the assumption
of geometric growth of the coefficients of the Adler
function difference is not the same as the assumption of
geometric growth of the coefficients of the correlator
difference itself.  The potential convergence problem, however,
may also be demonstrated to exist in the former case.}
\end{footnote}}
The contour improved analysis employing FESR's based on the
spectral weights thus has potentially significant
theoretical uncertainties.  

In light of the problems discussed above
for those FESR's based on the spectral
weights, $w_{L+T}^N$, our goal in the next section
will be to construct alternate weights which lead to FESR's
which bring these problems under control.

\section{The construction of alternate weight functions}

We begin our search for an alternate choice of weight function by
attempting to understand the source of the potential slow convergence 
of the contour-improved series noted above.  The goal will
be to find a weight such that, even were the unknown $g_k$, $k\geq 3$,
to grow geometrically, as assumed above, the 
tail of the contour-improved series would be small 
relative to the known terms, in contrast to the behavior shown
in Table I for 
the series corresponding to the spectral weights, $w_{L+T}^N$.
If we succeed in doing so, the reliability of the standard
approach, in which the truncation error is taken to
be given by the size of the last known term (in this case, $k=2$),
will, of course, be improved regardless of the actual
behavior of the unknown $g_k$.  We will then attempt to
simultaneously impose conditions which reduce the impact
of the experimental errors.

To study the source of the slow convergence of the contour-improved
series, it is useful to consider the behavior
of the factor $f_k(Q^2) \equiv m(Q^2)^2 a(Q^2)^k g_k$,
appearing in the integrand of $g_k A^{[w]}_k(s_0)$,
on the contour $\vert s\vert =s_0$.  Let
$w(y)$, $y=s/s_0$, be any analytic function real on the real $s$ axis,
and $Q^2=-s_0\exp (i\phi )$ ($\phi =0,\pi $ thus correspond
to timelike and spacelike points, respectively).  One then has
\begin{equation}
g_k\, A^{[w]}_k(s_0)={\frac{1}{\pi}}\int_0^\pi\, d\phi\, 
{\rm Re}\left[ f_k(Q^2) w\left( \exp ({i\phi})\right)\right]\ .
\label{fkdefn}
\end{equation}
The behavior of ${\rm Re}(f_k)$ and ${\rm Im}(f_k)$ as a function
of $\phi$, for $s_0=m_\tau^2$ and $k= 0,...,10$,
is shown in Figure 1.  We observe that both 
${\rm Re}(f_k)$ and ${\rm Im}(f_k)$
have zeroes on the circle $\vert s\vert =m_\tau^2$, 
and that these zeroes move with
the order $k$.  Moreover, while ${\rm Re} (f_k)$ 
(slowly) decreases with increasing $k$ for 
all angles $\phi$, the magnitude of ${\rm Im}(f_k)$ 
is sizeable in the region 
$\phi\geq \pi/2$ even for $k\geq 5$. 
This slow convergence in the backwards (spacelike) direction is
the origin of the slow convergence of the $k\geq 3$ tails of
the integrated series shown in Table I, since the factor
$(1-y)^{N+2}$ entering the weight $w_{L+T}^N$ has maximum
modulus at the spacelike point on the contour, and is more and more
sharply peaked in the backward direction as $N$ increases.  In addition,
the behavior of ${\rm Re}(f_2)$ and ${\rm Im}(f_2)$ happens
to be just such that, combined with the changes of sign of
the real and imaginary parts of $w_{L+T}^N$, there is a very
strong cancellation in the integral over $\phi$ (particularly
so for the case $N=0$).  This strong cancellation is the
origin of the ``accidental'' suppression of the magnitude
of the $k=2$ term.
As we have already seen in Table I, it is potentially
dangerous to use weights for which the integrals 
$A^{[w]}_k(s_0)$ are small for a particular $k$ (or for a small
number of values of $k$) {\it only due to such cancellations}. 
Higher order contributions can then easily be large again,
thereby spoiling the seemingly 
good convergence of the first few terms of the contour-improved series.

The behavior of the ${\rm Re}(f_k)$ and ${\rm Im}(f_k)$
displayed in Figure 1 allows one not only to understand
the origin of the potential convergence problem 
but also to construct alternate sum rules
which avoid it.  From Figure 1 it is evident that
convergence can be improved by avoiding weights which are
large in the spacelike direction.  The results of Ref.~\cite{kmfesr}
also indicate that, for the FESR framework
to be reliable at scales $\sim m_\tau^2$, 
it is necessary for the weight function
to have a zero at $s=s_0$ 
($y=1$).\begin{footnote}{Such a zero suppresses contributions from the OPE
representation in the region near the timelike real axis where, at scales
$\sim m_\tau^2$ and below, data shows that it breaks 
down~\cite{kmfesr}.}
\end{footnote}
We have found
two approaches useful for implementing
these constraints.
The first involves the use of polynomials
with ``shepherd'' zeros, i.e., zeros 
either on, or near, the regions of the contour
one wishes to suppress.  The second involves the construction of
weights, $w_p$, with ${\rm Im}(w_p)$ peaked on the contour at angles 
$\phi\leq \pi/2$, thereby avoiding large contributions from
${\rm Im}(f_k)$, $k>1$ (see Figure 1).
A convenient and effective choice is to take ${\rm Im}(w_p)$ to have
a Gaussian form on the contour.  Choosing the width of the Gaussian
to be $10^\circ$ and the center to be $\phi =\phi_p$, 
good convergence of the $k\geq 3$ tail of the 
integrated series can be obtained for any 
$20^\circ\leq \phi_p \leq 90^\circ$.  
Technically, these profiles 
can be well represented using polynomials of degree $K\approx 20$
\begin{equation}
w_p(y) = \sum_{i=0}^K a_i\, y^i.
\label{wpbasis}
\end{equation}
The coefficients $a_i$ are determined, upon normalizing ${\rm Im}(w_p)$
such that $w_p(0)=1$, by the Fourier integrals
\begin{equation}
a_0 = 1, \qquad a_k = {2\over \pi}\, \int_0^\pi d\phi\, {\rm Im}
\left( w_p(\phi) \right)\, \sin(k \phi), \qquad k=1\ldots K.
\label{Fouriercoeff}
\end{equation}

To summarize:  given the problems discussed above with those
FESR's involving the spectral weights, $w_{L+T}^N(y)$, we would
like to find, if possible, an alternate weight choice, $w(y)$,

(1) such that $w(y)$ is strongly suppressed in the region above
$s\sim 1\ {\rm GeV}^2$, in order to (a) reduce the degree
of cancellation between the $ud$ and $us$ spectral integrals,
(b) reduce the impact of the large experimental errors in the
$us$ spectral distribution above the $K^*$ region, and (c)
minimize the role of the longitudinal subtraction which must,
at present, be performed theoretically; and

(2) such that 
$w(y)$ emphasizes those regions of the contour $\vert s\vert =s_0$
for which the convergence of the $D=2$ series is favorable.

It is, of course, not {\it a priori} obvious that there exist 
$w(y)$ having the desired properties.  
We have, however, succeeded in constructing several
polynomial weights which 
do.\begin{footnote}{An important further 
restriction results from the observation that, in the FESR
framework, higher dimension contributions are suppressed
only by inverse powers of $s_0$; in order to
avoid generating potentially large, and unknown,
higher dimension contributions, therefore,
the coefficients of the polynomials we
construct should all be comparable in magnitude to the
leading coefficient, $a_0=1$.
We have chosen to implement this constraint by keeping all
coefficients less than $\sim 2$ in magnitude.}\end{footnote}
Since, as we will see below, the resulting weights do not contain
$w_{L+T}(y)$ as a factor, 
the approach is less inclusive than the analysis employing 
$w_{L+T}(y)$~\cite{pp98,pp99}, but it has the advantage of
being theoretically cleaner. 

The strategy involving shepherd zeros can be implemented with
the zeros either on or off the contour.  
The first weight we have constructed satisfying the criteria
above has all zeros on the contour, and is given by
\begin{equation}
w_{10}(y)=[1-y]^4[1+y]^2[1+y^2][1+ y+y^2]=
1-y-y^2+2y^5-y^8-y^9+y^{10}\ .
\label{oncontour}
\end{equation}
The absence of ${\cal O}(y^3,y^4)$ terms, which 
suppresses $D=8,10$ contributions, is an additional positive
feature of this weight.
The fourth order zero at $y=1$ and second order zero at
$y=-1$ provide the desired suppressions of the timelike
and spacelike regions.  An alternate family of weights
still having a fourth order zero at $y=1$,
but with the remaining zeros moved off the contour and at
a distance $r$ from the origin, is
\begin{equation}
\hat{w}(r,\cos \theta_1,\cos \theta_2,y)=[1-y]^4\left[ 1+{\frac{y}{r}}\right]^2
\left[ 1+2{\frac{y}{r}}\cos \theta_1+{\frac{y^2}{r^2}}\right]
\left[ 1+2{\frac{y}{r}}\cos \theta_2+{\frac{y^2}{r^2}}\right]
\label{offcontour}
\end{equation}
($\theta_1$ and $\theta_2$ give the angular positions 
of the pairs of off-contour complex conjugate zeros corresponding
to the last two factors, with respect to the spacelike direction).
The choice $(r,\cos \theta_1,\cos \theta_2)=(1.2,0.5,0.1)$
produces a second solution to the constraints above, one
whose biggest coefficient is $a_1=-4/3$. We
denote this solution by 
\begin{equation}
\hat{w}_{10}(y)=\hat{w}(1.2,0.5,0.1,y)\ .
\label{secondweight}
\end{equation}

In the approach based on weights which
have imaginary parts with a Gaussian profile 
on the contour, we choose a basis of such weights
having different centers, $\phi_p$.  As noted above,
so long as all the $\phi_p$ lie in
the interval $20^\circ\leq \phi_p\leq 90^\circ$,
all of the corresponding integrated $D=2$ perturbative series
will be under control.  We then form linear
combinations of these weights having different $\phi_p$
in such a way as to construct a new weight which not only retains
this good convergence, but at the same time has a zero of
sufficiently high order at $y=1$ to strongly suppress contributions
to the spectral integral from the region $y>0.5$.  The 
weight of this type which most successfully
satisfies the criteria discussed above 
has a rapid high-$s$ falloff produced by a
$6^{th}$ order zero 
at $y=1$, a largest coefficient $a_4=2.087$, and is given by
\begin{eqnarray}
w_{20}(y)= (1-y)^6 \big[ 1 &&+4.2451 y +9.4682 y^2
+14.4155 y^3+16.4589 y^4+14.6598 y^5\nonumber \\
&& +10.2818 y^6+5.5567 y^7
+2.1157 y^8+0.3520 y^9-0.2065 y^{10} \nonumber\\
&& -0.2154 y^{11}-0.1040 y^{12}-0.03040 y^{13}-0.0045 y^{14} \big] \ .
\label{w20explicit}
\end{eqnarray}

The (vastly) improved convergence of the $k\geq 3$
tail of the integrated
$D=2$ series for the weights $w_{10}$, $\hat{w}_{10}$
and $w_{20}$ is displayed in Table II.
The entries, as in Table I, have been rescaled by the corresponding
$k=0$ value, and hence
correspond to the ratios, 
$g_kA^{[w]}_k(m_\tau^2)/A^{[w]}_0$.  
The results also show that an estimate of the truncation
error given by the magnitude of the $k=2$ term is, for the new weights,
almost certainly a very conservative one.  
We will demonstrate, in the next section, that the suppression
of the high-$s$ region of the spectrum produced by the
new weights is also sufficient to significantly reduce
the impact of the experimental errors.

\section{Numerical Analysis and Results}
In performing the numerical analysis of the FESR's constructed
above, we employ the ALEPH data for the nonstrange and 
strange number distributions\begin{footnote}{The 
1998 tabulation of the nonstrange data 
receives a small overall normalization correction as a result
of the shift in $R^{us}_\tau$ between the preliminary 1998 and final
1999 analyses.  We thank Shaomin Chen for bringing this point
to our attention.}
\end{footnote} and PDG98 values for $f_K$, $f_\pi$, $\vert V_{ud}\vert$ and
$\vert V_{us}\vert$.
As noted above, the weights have been chosen in such a way that,
although theoretical input is required in order to subtract
the longitudinal contributions to the experimental number
distributions, and hence obtain the $L+T$ spectral
functions, the effect of this subtraction on the final
value of $m_s$ is negligible.  We will quantify this
statement below.  
Once the $L+T$ spectral function
has been determined, it is a straightforward matter
to evaluate the weighted $L+T$ spectral integrals.
The choice of steeply falling weights ensures that
the strange spectral integrals are dominated by the
$K$ and $K^*$ contributions, for which the experimental
errors are much smaller than those of the rest of the
strange number distribution.  This plays a major role
in reducing the impact of experimental errors on the
final extracted value of $m_s$.
To get a realistic determination of these errors
it is important to separate correlated
and uncorrelated errors, and also to take into account
the strong correlations between the spectral integrals
involving different weights.

The nature of the longitudinal subtraction differs significantly
in the low-$s$ and high-$s$ ($>\sim 1\ {\rm GeV}$)
regions.  For low $s$, the $\pi$ and $K$ pole subtractions are experimentally
unambiguous.  For high $s$ (the resonance region), the longitudinal
contributions are proportional to 
$(m_s\pm m_u)^2$, $(m_d\pm m_u)^2$, 
for $us$, $ud$, respectively, and hence
dominated by the $us$ contributions.  The longitudinal
$us$ vector contribution is inferred from the strange scalar
spectral function of Ref.~\cite{cfnp}.  This procedure is consistent
provided the value of $m_s$ resulting from the
present analysis is compatible with that from the strange
scalar channel~\cite{kmss}, which it
turns out to be.  The longitudinal $us$ axial vector
contribution is similarly inferred from the 
spectral function of the strange pseudoscalar channel.  The latter 
is obtained by fixing
the excited resonance decay constants 
of a sum-of-resonances spectral ansatz through
matching of the hadronic and OPE sides of a 
family of ``pinch-weighted'' FESR's, in analogy to the
analysis of Ref~\cite{kma0}.\begin{footnote}{The 
corresponding procedure works very well
in the isovector vector channel, where the results can be
checked against the well-known experimental
spectral function~\cite{kma0}.  A similar statement
is true even in channels with strongly attractive
interactions near threshold, for which the spectral
function will be poorly represented {\it near threshold}
by the tail of a Breit-Wigner resonance form with ``conventional''
$s$-dependent width.  For example, using the value of
$m_s$ obtained from the strange scalar channel analysis
as input and redoing the strange scalar channel analysis,
using now a sum-of-resonances spectral ansatz in place of
the more realistic ansatz of Ref.~\cite{cfnp}, one
finds that the ansatz of Ref.~\cite{cfnp} is 
well-reproduced in the region of the dominant $K_0^*(1430)$
peak.  One can also use this approach to check the
self-consistency between the assumed longitudinal
contributions and the output $m_s$ value in kinematic-weight-based
analysis of Ref.~\cite{cdh,ALEPHstrange}.  It turns out that the 
high-$s$ longitudinal contributions assumed are more
than a factor of $2$ smaller than would be expected
based on the extracted value of $m_s$.  If one employs
the PDG98 values for $\vert V_{us}\vert$ and $f_K$,
as discussed above, however, the assumed longitudinal
contribution becomes compatible within the errors
assigned to it in Ref.~\cite{cdh,ALEPHstrange}.}\end{footnote}
The input value of $m_s$ required for this analysis should,
in principle, be determined iteratively. 
We have, however, employed as input
the value of $m_s$ obtained from the strange scalar
analysis of Ref.~\cite{kmss}, $m_s(1 \ {\rm GeV})=159\pm 11 \ {\rm MeV}$.
This turns out to be consistent with our final result for $m_s$.
Moreover, for the steeply-falling 
weights employed in our analysis, the sum of the high-$s$
$V$ and $A$ longitudinal subtractions is at the $<0.1\%$ level
of the $us$ spectral integral, and hence at the $<1\%$
level in the $ud$-$us$ difference.  As such, even were
our evaluation to be in error by $100\%$, the effect
on $m_s$ would be completely negligible on the scale of the other
errors present in the analysis.

On the OPE side, we retain contributions up to and including
$D=8$.  The leading $D=2$ term was given above.

The $D=4$ contribution is~\cite{bnp,ck93}
\begin{eqnarray}
\left[ \Pi (Q^2)\right]_{(D=4)}&=&
{\frac{2}{Q^4}}\Bigl[ \left( m_\ell <\bar{\ell}\ell >-I_s\right)
\left( 1-a(Q^2)-{\frac{13}{3}}a(Q^2)^2\right)\nonumber \\
&&\qquad+{\frac{3}{7\pi^2}}m_s^4(Q^2)\left( {\frac{1}{a(Q^2)}}
-{\frac{7}{12}}\right)\Bigr]\, ,
\label{fbd4}
\end{eqnarray}
where $I_s$ is the usual RG invariant modification of the
non-normal-order strange quark condensate~\cite{cs88}, 
$m_\ell$ is the average of
the light $u$, $d$ masses, and $<\bar{\ell}\ell >$ is
the light ($u,d$) condensate.  We use 
the quark mass ratios determined from the ChPT analyses of
Ref.~\cite{leutwylermq}, 
the GMO relation
$2m_\ell <\bar{\ell}\ell >=-f_\pi^2 m_\pi^2$, 
and the range of values
$0.7< \langle \bar{s}s\rangle /\langle \bar{\ell}\ell \rangle <1$~\cite{jm,cps}
for the ratio of condensates.  The contour integrals are performed
as described below.

For the $D=6$ contribution
we employ a rescaled version of the vacuum saturation approximation
(VSA).  From the results of Ref.~\cite{bnp}, one finds
\begin{equation}
\left[ \Pi (Q^2)\right]_{(D=6)}=
{\frac{64\pi\rho\alpha_s}{81Q^6}}\Bigl[ <\bar{\ell}\ell >^2
-<\bar{s}s>^2\Bigr]\, ,
\label{fbd6}
\end{equation}
where $\rho$ represents a multiplicative rescaling of
the VSA estimate.
The analogous rescaling has been determined
empirically for the isovector vector channel and the 
isospin-breaking vector $38$ correlator, and found to be $\sim 5$
in both cases~\cite{narisonrho,kmcew98}.  
For the weights
employed in our analysis, it turns out that the integrated
$D=6$ contributions are very small.  We are, therefore, able to
employ the very conservative estimate $\rho =5\pm 5$ for
the degree of VSA violation without significantly affecting
the overall theoretical error.  The combination 
$\rho\alpha_s <\bar{q}q>^2$ in Eq.~(\ref{fbd6})
is to be understood as an effective RG-invariant combination
for the evaluation of the OPE contour integrals.

Finally, for the $D=8$ contribution, we assume
\begin{equation}
\left[ \Pi (Q^2)\right]_{(D=8)}=
{\frac{C_8}{Q^8}}\, .
\label{fbd8}
\end{equation}
For $w_{10}$ this term does not contribute to the integrated OPE;
for $w_{20}$ and $\hat{w}_{10}$,
the value of the effective RG-invariant condensate
combination, $C_8$, is to be determined as part of the analysis.

As noted above, the OPE contour integrals 
(for all $D$) are performed using the contour improvement
prescription.  Four-loop versions of the running mass and coupling
are employed.  To be specific, we have solved 
analytically for the running mass and coupling
using the 4-loop truncated versions of the $\beta$~\cite{beta4} and
$\gamma$~\cite{gamma4} functions, with the value
determined in nonstrange hadronic $\tau$ decays,
$\alpha_s (m_\tau^2)=0.334\pm 0.022$~\cite{ALEPHlight},
as input.  Following conventional practice,
we take the error associated with the truncation of 
the perturbative series for the Wilson coefficient of
the $D=2$ term at ${\cal O}(a^2)$ to be equal to the value of
the last (${\cal O}(a^2)$) contribution retained.  In light of the discussion
above we consider this to represent an extremely conservative
estimate.  

From the point of view of uncertainties on the OPE side,
the $w_{10}$ sum rule is favored over the $\hat{w}_{10}$
and $w_{20}$ sum rules for three reasons: (1) it has no $D=8,10$ contributions,
(2) it has the smallest truncation error, and (3) it has the smallest
errors associated with uncertainties in the input values of
the $D=4$ and $D=6$ condensates.\begin{footnote}{Combining 
the errors associated
with truncation, the condensate input values, and the uncertainty
on $\alpha_s(m_\tau^2)$ in quadrature, the resulting errors
on $m_s$ are $7.7\%$, $8.2\%$ and $8.4\%$ for $w_{10}$,
$\hat{w}_{10}$ and $w_{20}$, respectively.}\end{footnote}
In Table III we display, as a function of $s_0$, the
extracted values of $m_s(1 \ {\rm GeV}^2)$
obtained from the $w_{10}$ sum rule, analyzed
neglecting contributions of dimension 12
and higher.  Central values have been used for all input
on the OPE side and for the experimental spectral data.
For the analysis to be self-consistent, the extracted value of
$m_s$ should be independent of $s_0$.  This will be
true for $s_0$ sufficiently large that the $D\geq 12$
contributions are negligible.  As $s_0$ is decreased,
the extracted $m_s$ values should eventually deviate from a constant,
signalling the growth
of the higher dimension terms.  From the
Table we see that the range $2.75\ {\rm GeV}^2
< s_0<3.15\ {\rm GeV}^2$ provides an extremely good
window of stability.  In view of the falloff begining around
$s_0\sim 2.55\ {\rm GeV}^2$, we will work in the range $s_0\geq 2.55
\ {\rm GeV}^2$ in the discussions which follow.
It is worth stressing that the central values obtained 
from $w_{20}$ and $\hat{w}_{10}$ sum rules, though having
slightly larger theoretical errors, are nonetheless completely
consistent with those above:  in the window
$2.55\ {\rm GeV}^2\leq s_0\leq 3.15\ {\rm GeV}^2$, 
one finds that the range of solutions for $m_s(1\ {\rm GeV}^2)$ 
lies between $156$ and $161 \ {\rm MeV}$ for $w_{20}$, 
$158$ and $164 \ {\rm MeV}$ for $\hat{w}_{10}$, and,
as we saw already in Table III, $159$ and $163\ {\rm MeV}$ for 
$w_{10}$.  In contrast, the $w_{L+T}$ sum rule, for which
the longitudinal subtraction is important, and the
$D=2$ convergence is not well under control, yields
a range between $161$ and $184$ (with, moreover, inconsistent
solutions for $C_8$).

From the point of view of the impact of the errors present
in existing experimental data,
the theoretically favored
$w_{10}$ weight is, unfortunately, no longer the favored one.
The reason is that, although the 
impact of the errors in the high-$s$ region of the
$us$ spectrum has been strongly suppressed by the 
rapid falloff of the weights employed, the $ud$-$us$ 
cancellation is still rather close ({\it e.g.}, at $s_0=m_\tau^2$,
to the level of $6.0\%$
for $w_{10}$, $6.8\%$ for $\hat{w}_{10}$ and $8.6\%$
for $w_{20}$, to be compared with $3.7\%$, $6.5\%$
and $9.3\%$ for the $w_{L+T}^N$, $N=0,1,2$.)  Although the
dominant errors (those from the $K^*$ region of the
$us$ spectrum) are reasonably small, they are still large
enough that the {\it relative} size of the residual statistical error
grows very rapidly with the increase in the degree
of cancellation.  Thus, {\it e.g.}, at $s_0=m_\tau^2$,
the statistical error represents
$42\%$, $36\%$, $26\%$,
$77\%$, $38\%$ and $23\%$ of the $ud$-$us$ spectral difference 
for the $w_{10}$, $\hat{w}_{10}$,
$w_{20}$, $w_{L+T}^0$, $w_{L+T}^1$, and $w_{L+T}^2$ 
sum rules, respectively.\begin{footnote}{Because of the high
degree of cancellation, reducing $s_0$, which increases
the degree of suppression of the (already small) high-$s$
$us$ contributions, still has a non-trivial effect; {\it e.g.},
the relative statistical error for the $w_{20}$ sum rule
is reduced from $26\%$ to $19\%$ when $s_0$ is
lowered from $m_\tau^2$ to $2.55\ {\rm GeV}^2$.}\end{footnote}
The present experimental situation is, therefore, such that
the errors on our final result for $m_s$ are
minimized by working with $w_{20}$, rather than $w_{10}$.

Working with the $w_{20}$ sum rule in the window specified
above we find, for our best fit,
\begin{equation}
m_s(1\ {\rm GeV}^2)= 158.6\pm 18.7\pm 16.3\pm 13.3\ {\rm MeV}\ ,
\label{finalms}
\end{equation}
which is equivalent to
\begin{equation}
m_s(4\ {\rm GeV}^2)=115.1\pm 13.6\pm 11.8\pm 9.7\ {\rm MeV}\ ,
\label{otherfinalms}
\end{equation}
where in both of Eqs.~(\ref{finalms})
and (\ref{otherfinalms}) the first error is statistical, the second is due
to the uncertainty on $\vert V_{us}\vert$, and the third
theoretical.  The theoretical error has been obtained by
combining the following in
quadrature (where we quote the numerical values
corresponding to Eq.~(\ref{finalms}) to be specific):  
$\pm 5.2\ {\rm MeV}$, associated with the error on
$\alpha_s (m_\tau^2)$; $\pm 3.6\ {\rm MeV}$, associated with the
uncertainty in $<\bar{s}s>/<\bar{\ell}\ell >$;
$\pm 1.6\ {\rm MeV}$, associated with the variation of 
$m_s$ within the window $2.55\ {\rm GeV}^2\leq s_0\leq
m_\tau^2$; $\pm 0.6\ {\rm MeV}$, associated with the uncertainty in the
VSA-violating parameter, $\rho$; and $\pm 11.6\ {\rm MeV}$,
associated with truncation of the $D=2$ series.  The
latter obviously remains the dominant 
source of theoretical error, despite 
the significant improvement produced by the use of the new
weights.  Figure 2 displays the quality of the match between
the OPE and spectral integral sides of the $w_{20}$ sum rule
corresponding to the fit above; the agreement in the
previously-established stability window, $s_0>2.55\ {\rm GeV}^2$,
is obviously
excellent.  The divergence of the OPE and spectral integral
curves below $s_0\sim 2.55\ {\rm GeV}^2$ is precisely
what one would expect based on the observation above that,
for the $w_{10}$ sum rule, $D>10$ contributions, not included
in the truncated OPE representation, begin to become important in
this region.

The result of Eqs.~(\ref{finalms})
and (\ref{otherfinalms}) is
in good agreement with the strange
scalar channel results of Refs.~\cite{cfnp} and ~\cite{kmss},
the strange pseudoscalar channel result of Ref.~\cite{dps},
and the recent hadronic $\tau$ decay analysis of
Ref.~\cite{pp99}, but, we believe, has signficantly
reduced theoretical and experimental errors.
In particular, the statistical error has, at this point,
been reduced almost to the level of that associated with
the uncertainty in $\vert V_{us}\vert$.

Improvements in the accuracy of the experimental $us$
spectral data, in particular in the $K^*$ region, could
lead to a significant improvement in the size of the
statistical error.  Such an improvement should be possible
using BaBar data\cite{roney}.  Reduced uncertainties in
our knowledge of $\vert V_{us}\vert$ would also be helpful.
On the theoretical side, while significant improvements
in the accuracy of the spectral data would allow one to
move from the $w_{20}$ to the $w_{10}$ sum rule,
the decrease in the theoretical uncertainty that would
result from this shift would
be only $\sim 1.3\ {\rm MeV}$.  Far more likely to lead to
a significant improvement in the 
size of the theoretical error would
be a computation of the ${\cal O}(a^3)$ coefficient in
the $D=2$ contribution to the flavor-breaking
correlator difference, $\Pi$.

\acknowledgements
{The authors would like to thank A. H\"ocker and S. Chen
for providing detailed information on the ALEPH
nonstrange and strange spectral distributions, S. Chen
for pointing out the normalization correction to the 1998
nonstrange data necessitated by the results of the 1999
strange data analysis, and G. Colangelo for his
collaboration at an early stage of this work.  KM acknowledges
the ongoing support of the Natural Sciences and
Engineering Research Council of Canada, and the hospitality of the
Special Research Centre for the Subatomic Structure of Matter at the
University of Adelaide, where much of this work was performed, and
JK the partial support of the Schweizerischer Nationalfonds and
the EEC-TMR program, Contract No. CT 98-0169.}

\noindent
\begin{table}
\caption{OPE convergence of the ``contour improved'' $D=2$
contributions, $g_k A^{[w_{L+T}^N]}_k(m_\tau^2)$, as a function
of the contour improved order, $k$,
for the spectral weights, $w_{L+T}^N(y)=(1-y)^{N+2}(1+2y)$,
assuming geometric growth of coefficients
beyond ${\cal O}(\alpha_s^2)$.  All entries have been rescaled
by the corresponding entry for $k=0$.}
\begin{tabular}{cccccccccccc}
Weight&$k=0$&$k=1$&$k=2$&$k=3$&$k=4$&$k=5$&$k=6$&$k=7$&$k=8$&$k=9$&$k=10$ \\
\hline
$w_{L+T}^0$&1&0.143&-0.007&-0.145&-0.237&-0.286&-0.294&-0.272&-0.233&
-0.187&-0.141 \\
$w_{L+T}^1$&1&0.209&0.100&-0.027&-0.143&-0.232&-0.287&-0.308&-0.300&
-0.272&-0.233 \\
$w_{L+T}^2$&1&0.257&0.187&0.076&-0.048&-0.143&-0.260&-0.324&-0.357&
-0.359&-0.339 \\
\end{tabular}\label{table1}
\end{table}

\noindent
\begin{table}
\caption{OPE convergence of the ``contour improved'' $D=2$
contributions, $g_k A^{[w]}_k(m_\tau^2)$, as a function
of the contour improved order, $k$,
for the weights, $w_{10}$, $\hat{w}_{10}$, and $w_{20}$,
assuming geometric growth of coefficients
beyond ${\cal O}(\alpha_s^2)$.  All entries have been rescaled
by the corresponding entry for $k=0$.}
\begin{tabular}{cccccccccccc}
Weight&$k=0$&$k=1$&$k=2$&$k=3$&$k=4$&$k=5$&$k=6$&$k=7$&$k=8$&$k=9$&$k=10$ \\
\hline
$w_{20}$&1&0.262&0.213&0.143&0.073&0.018&-0.017&-0.033&-0.034&
-0.027&-0.016 \\
$w_{10}$&1&0.232&0.165&0.092&0.032&-0.008&-0.030&-0.038&-0.038&
-0.035&-0.032 \\
$\hat{w}_{10}$&1&0.248&0.193&0.125&0.064&0.019&-0.009&-0.023&-0.026&
-0.024&-0.020 \\
\end{tabular}\label{table2}
\end{table}

\noindent
\begin{table}
\caption{The extracted value of $m_s (1\ {\rm GeV}^2)$ in MeV as
a function of $s_0$ for the weight $w_{10}$ having no $D=8,10$
contributions.}
\begin{tabular}{lccccc}
$s_0$ (GeV$^2$):&2.35&2.55&2.75&2.95&3.15 \\
$m_s (1\ {\rm GeV}^2)$ (MeV):&153.2&159.0&162.2&163.4&163.2 \\
\end{tabular}\label{table3}
\end{table}

\noindent
\begin{figure} [htb]
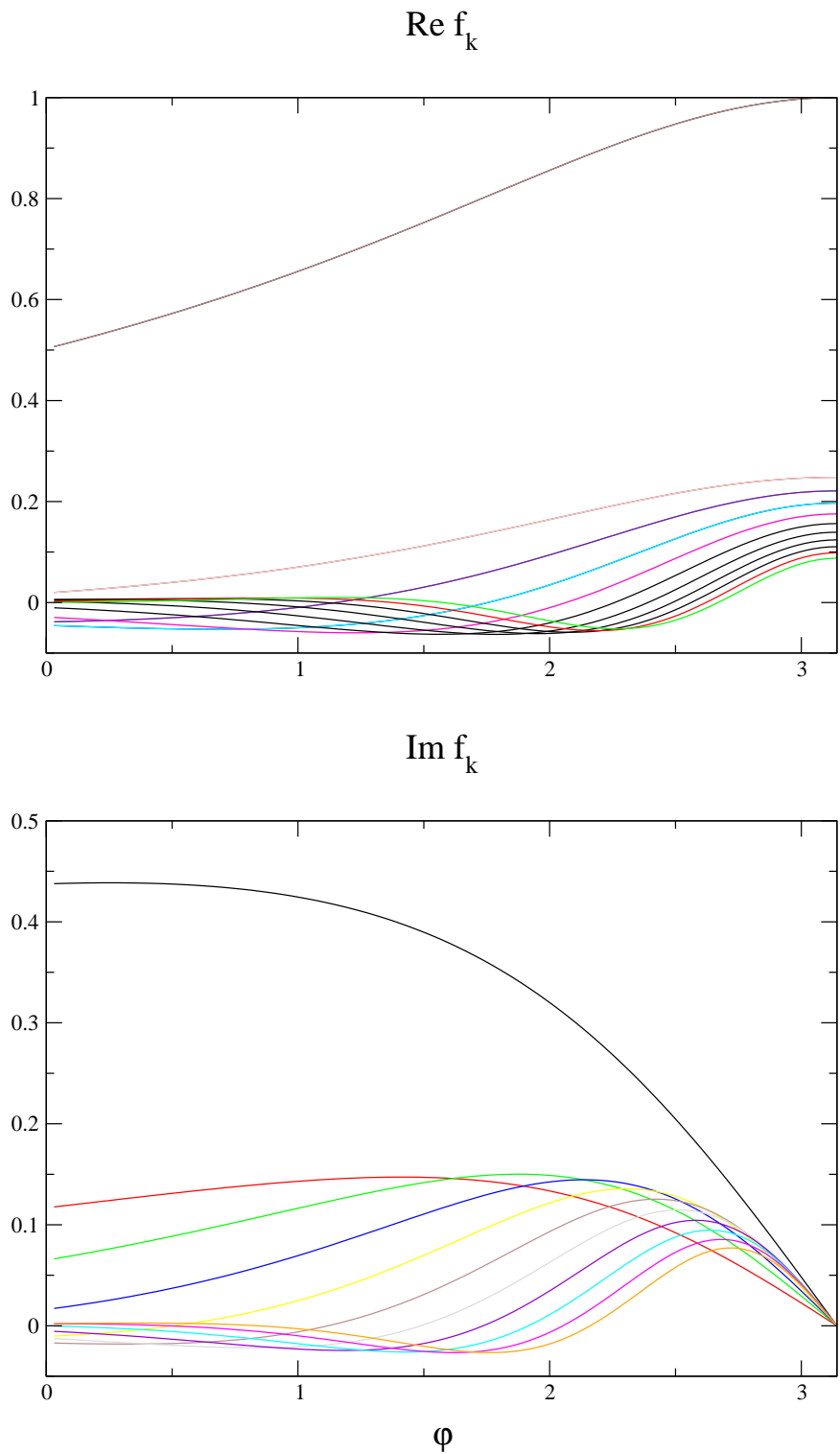

\centering{\
\psfig{angle=-90,figure=Refk_new.eps,height=10.0cm}}
%\qquad\qquad\qquad&

\centering{\ 
\psfig{angle=-90,figure=Imfk_new.eps,height=10.0cm}}
%\end{tabular}}
%\psfig{figure=ReImfk.eps,height=5.cm}}
\vskip .3in
\caption{The real and imaginary parts of $f_k$, $k=0,\cdots ,10$,
at scale $m_\tau^2$,
where $k$ labels the power of $\alpha_s$. $f_k$ is defined
explicitly in the text.}
\label{figReImfk}
\end{figure}
\vfill\eject

\ 

\vskip .3in\noindent
\begin{figure} [htb]
\centering{\
\psfig{figure=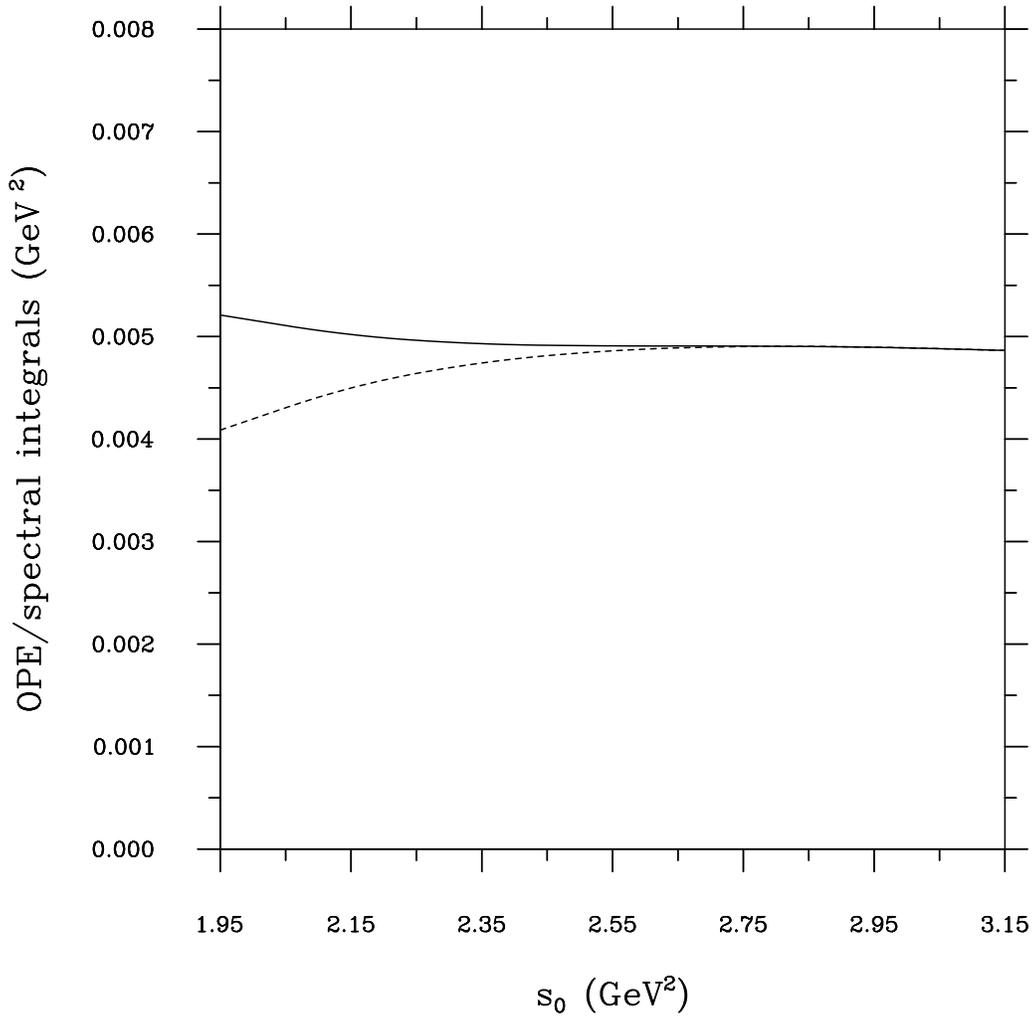,height=15.cm}}
\vskip .3in
\caption{The agreement between the OPE and hadronic sides of
the FESR corresponding to the weight, $w_{20}(y)$ for
$1.95\ {\rm GeV}^2\leq s_0\leq m_\tau^2$.
The solid line is the OPE side, using
the values of $m_s$ and $C_8$ obtained in the fitting procedure
described in the text.
The dashed line is the
hadronic side, obtained using the ALEPH spectral data from which
the longitudinal component has been subtracted as described
in the text. }
\label{figtwo}
\end{figure}

\vfill\eject

\vfill\eject

\begin{references}
\bibitem{bpr}J. Bijnens, J. Prades and E. de Rafael, Phys. Lett.
{\bf B348}, 226 (1995).
\bibitem{jm}M. Jamin and M. M\"unz, Z. Phys. {\bf C66}, 633 (1995);
M. Jamin, Nucl. Phys. B (Proc. Suppl.) {\bf 64}, 250 (1998).
\bibitem{cps}K.G. Chetyrkin, D. Pirjol and K. Schilcher, 
Phys. Lett. {\bf B404}, 337 (1997).
\bibitem{n38}S. Narison, Phys. Lett. {\bf B358}, 113 (1995).
\bibitem{cfnp}P. Colangelo, F. De Fazio, G. Nardulli and N. Paver, 
Phys. Lett. {\bf B408}, 340 (1997).
\bibitem{bgm}T. Bhattacharya, R. Gupta and K. Maltman, Phys.
Rev. {\bf D57}, 5455 (1998).
\bibitem{km38}K. Maltman, Phys. Lett. {\bf B428}, 179 (1998).
\bibitem{dps}C. Dominguez, L. Pirovano and K. Schilcher, Phys. Lett.
{\bf B425}, 193 (1998).
\bibitem{kmss} K. Maltman, Phys. Lett. {\bf B462}, 195 (1999).
\bibitem{ck93}K.G. Chetyrkin and A. Kwiatkowski, Z. Phys. {\bf C59}, 
525 (1993) and hep-ph/9805232.
\bibitem{kmtpr}K. Maltman, Phys. Rev. {\bf D58:} 093015 (1998).
\bibitem{pp98}A. Pich and J. Prades, JHEP {\bf 9806:013} (1998).
%\bibitem{ckp98}K.G. Chetyrkin, J.H. Kuhn, A.A. Pivovarov,
%Nucl. Phys. {\bf B533}, 473 (1998).
\bibitem{cdh}S. Chen, M. Davier and A. Hocker, LAL-98-90, Nov. 1998.
\bibitem{pp99}A. Pich and J. Prades, JHEP {\bf 9910:004} (1999).
\bibitem{ALEPHstrange}The ALEPH Collaboration, Eur. Phys. J. {\bf C11},
599 (1999), hep-ex/9903015.
\bibitem{n3s}S. Narison, Phys. Lett. {\bf B466}, 345 (1999); hep-ph/9905264.
\bibitem{nrev}S. Narison, Nucl. Phys. Proc. Suppl. {\bf 86},
242 (2000), hep-ph/9911454.
\bibitem{lat1}R. Gupta and T. Bhattacharya, Phys. Rev. {\bf D55}, 7203
(1997) and Nucl. Phys. (Proc. Suppl.) {\bf 63}, 45 (1998);
B.J. Gough {\it et al.}, Phys. Rev. Lett. {\bf 79}, 1662 (1997).
\bibitem{lat2}R.D. Kenway, Nucl. Phys. (Proc. Suppl.) {\bf 73}, 16 (1999).
\bibitem{lat3}V. Lubicz, Nucl. Phys. (Proc. Suppl.) {\bf 74}, 291 (1999).
\bibitem{latticerecent}T. Blum, A. Soni and M. Wingate, Phys. Rev. {\bf D60:} 
114507 (1999);
J. Gardner, J. Heitger, R. Sommer and H. Wettig
(ALPHA/UKQCD), Nucl. Phys. {\bf B571}, 237 (2000);
S. Aoki, {\it et al.} (JLQCD), Phys. Rev. 
Lett. {\bf 82}, 4392 (1999);
S. Aoki {\it et al.} (CP-PACS), Phys. Rev.
Lett. {\bf 84}, 238 (2000);
W. G\"ockeler {\it et al.} (QCDSF), hep-lat/9908005;
D. Becirevic, V. Lubicz, G. Martinelli
and M. Testa, hep-lat/9909039;
A. Ali Khan, {\it et al.} (CP-PACS), 
hep-lat/9909050 and hep-lat/0004010.
\bibitem{kmcew98}K. Maltman and C.E. Wolfe, Phys. Rev. {\bf D59:} 
096003 (1999).
\bibitem{ALEPHlight}R. Barate {\it et al.} (The ALEPH Collaboration),
Z. Phys. {\bf C76}, 379 (1997); Eur. Phys. J. {\bf C4}, 409 (1998).
\bibitem{gk96}E. Golowich and J. Kambor, Phys. Rev. {\bf D53}, 2651
(1996). 
\bibitem{dk99}S. D\"urr and J. Kambor, Phys. Rev. {\bf D61:} 114025 (2000).
\bibitem{pdg98}Review of Particle Properties, Eur. Phys. J. {\bf C3}, 1 (1998).
\bibitem{tau1}Y.S. Tsai, Phys. Rev. {\bf D4}, 2821 (1971);
H.B. Thacker and J.J. Sakurai, Phys. Lett. {\bf B36}, 103 (1971);
F.J. Gilman and D.H. Miller, Phys. Rev. {\bf D17}, 1846 (1978);
F.J. Gilman and S.H. Rhie, Phys. Rev. {\bf D31}, 1066 (1985).
\bibitem{tau2}E. Braaten, Phys. Rev. Lett. {\bf 60}, 1606 (1988);
S. Narison and A. Pich, Phys. Lett. {\bf B211}, 183 (1988); E. Braaten, Phys.
Rev. {\bf D39}, 1458 (1989); S. Narison and A. Pich, Phys. Lett. {\bf B304},
359 (1993).
\bibitem{bnp}E. Braaten, S. Narison and A. Pich, Nucl. Phys. {\bf B373},
581 (1992).
\bibitem{ledp92}A.A. Pivovarov, Sov. J. Nucl. Phys. {\bf 54}, 676 (1991)
and Z. Phys. {\bf C53}, 461 (1992);
F. Le Diberder and A. Pich, Phys. Lett. {\bf B286}, 147
(1992) and {\bf B289}, 165 (1992).
\bibitem{pichrev}A. Pich, hep-ph/9704453, in ``Heavy Flavors II'',
eds. A.J. Buras and M. Lindner, World Scientific, 1997.
\bibitem{ms88}W.J. Marciano and A. Sirlin, Phys. Rev. Lett. {\bf 61},
1815 (1988).
\bibitem{kmfesr}K. Maltman, Phys. Lett. {\bf B440}, 367 (1998).
\bibitem{kma0}K. Maltman, Phys. Lett. {\bf B462}, 14 (1999).
\bibitem{cs88}K. Chetyrkin and K.G. Spiridonov, Sov. J. Nucl.
Phys. {\bf 47}, 3 (1988).
\bibitem{leutwylermq}H. Leutwyler, Phys. Lett. {\bf B374}, 163 (1996);
Phys. Lett. {\bf B378}, 313 (1996) and hep-ph/9609467.
\bibitem{narisonrho}S. Narison, Phys. Lett. {\bf B361}, 121 (1995).
\bibitem{beta4}T. van Ritbergen, J.A.M. Vermaseren and S.A. Larin,
Phys. Lett. {\bf B400}, 379 (1997).
\bibitem{gamma4}K.G. Chetyrkin, Phys. Lett. {\bf B404}, 161 (1997);
T. Van Ritbergen, J.A.M. Vermaseren and S.A. Larin,
Phys. Lett. {\bf B405}, 327 (1997).
\bibitem{roney}M. Roney, private communication.
\end{references}
\end{document}